\documentclass[twocolumn,secnumarabic,amssymb, nobibnotes, aps, prd]{revtex4-1}

\usepackage[caption=false]{subfig}
\usepackage{amssymb}
\usepackage{amsmath}
\usepackage{commath}
\usepackage{graphicx,bm}
\usepackage{verbatim}
\usepackage{graphicx}
\usepackage{dcolumn}
\usepackage{bm}
\usepackage{multirow}
\usepackage{booktabs}
\usepackage{chngpage}
\usepackage{float}
\usepackage{setspace}

\newcommand{\bef}{\begin{figure}}
\newcommand{\eef}{\end{figure}}
\newcommand{\bec}{\begin{center}}
\newcommand{\eec}{\end{center}}
\newcommand{\be}{\begin{equation}}
\newcommand{\ee}{\end{equation}}

\begin{document}

\title{Implications on the cosmic coincidence by a dynamical extrinsic curvature}%

\author{A.J.S Capistrano}%
\email[]{abraao.capistrano@unila.edu.br}
\affiliation{1 Federal University of Latin-American Integration, 85867-970, P.o.b: 2123, Foz do Igua\c{c}u-PR, Brazil}
\affiliation{2 Casimiro Montenegro Filho Astronomy Center, Itaipu Technological Park, 85867-900, Foz do Iguassu-PR, Brazil}

\author{L.A. Cabral}%
\email[]{cabral@uft.edu.br}
\affiliation{3 Federal University of Tocantins, 77804-970, Aragua\'{i}na-TO, Brazil}

\date{\today}

\begin{abstract}
In this work, we apply the smooth deformation concept in order to obtain a modification of Friedmann equations. It is shown that the cosmic coincidence can be at least alleviated using the dynamical properties of the extrinsic curvature. We investigate the transition from nucleosynthesis to the coincidence era obtaining a very small variation of the ratio $r=\frac{\rho_{m}}{\rho_{ext}}$, that compares the matter energy density to extrinsic energy density,  compatible with the known behavior of the deceleration parameter. We also show that the calculated ``equivalence'' redshift matches the transition redshift from a deceleration to accelerated phase and the coincidence ceases to be. The dynamics on $r$ is also studied based on Hubble parameter observations as the latest Baryons Acoustic Oscillations/Cosmic Microwave Background Radiation (BAO/CMBR) + SNIa.
\end{abstract}

\maketitle

\section{Introduction}
Modifications of gravity at very large scales have been predominantly associated  with  the existence of extra  dimensions, an idea that has been  explored   in various  theories beyond the standard model of particle physics. As proposed  in \cite{ADD},  these extra dimensions may also provide a possible explanation for the huge difference between the two fundamental energy scales in nature, namely, the electroweak and Planck scales [$M_{Pl}/m_{EW} \sim 10^{16}$].  Essentially,
if our four-dimensional  space-time is embedded in a higher dimensional space, then  the  gravitational field  can also propagate  along the  extra dimensions,  while the  standard  gauge interactions remain  confined  to the four-dimensional spacetime. The impact of  such program  in theoretical and observational cosmology  has  been substantial  and  discussed  at  length as,  e.g.,  in Refs.~\cite{RS,RS1,7,Tsujikawa,9,maartens,maeda,sepangi,sepangi1,QBW,rostami,jalal,heydari} with or without junction conditions. However, the  majority of  contributions  on  this  theme are   string  (or brane)  inspired  models,  in  which the brane-world is  generated  by the  motion of  a three-dimensional brane  in the bulk.

In a different approach, we have explored the physical applications and implications of using the concept of smooth deformations of Nash's  theorem \cite{Nash} where the  bulk geometry is  defined   by  the   Einstein-Hilbert principle of smooth  curvature and only gravity  necessarily  propagates in the  extra  dimensions.  In that  respect, it  should  be  emphasized that the  four-dimensionality of the embedded  space-time is regarded  as   a  consequence of   the  structure of the gauge  field  equations, which is applicable to  all   embedded  space-times  (and  not just to  a  fixed boundary). Such  extended  notion of  confinement  is   consistent  with  the Einstein-Hilbert  dynamics for the bulk  geometry,  and  also to  the extrinsic  curvature satisfying the Gupta  equations \cite{Gupta,GDEI} for spin-2  fields in  space-time.

In what follows, we focus on the coincidence problem, sometimes referred as the ``new'' cosmological constant (CC) problem \cite{weinberg,weinberg2, shapiro1,shapiro2,shapiro3} which, in short, refers to the lack of a proper explanation of the present-day contributions of vacuum energy density $\Omega^{0}_{\Lambda}$ and the matter energy density $\Omega^{0}_{m}$ that are roughly the same order, i.e, $\Omega^{0}_{m}\sim \Omega^{0}_{\Lambda}$ since CC is considered the main cause of the accelerated expansion of the universe. Hence, as commonly stated, it seems we are living is a very special phase to observe it.

In this work, we study a process to at least alleviate the cosmic coincidence through the introduction of the dynamics of the extrinsic curvature $k_{\mu\nu}$, interpreted as an additional component to the gravitational field. Then, CC is totally uncorrelated to this process and is not considered here. We also analyze the necessary conditions of our model in order to do not jeopardize the nucleosynthesis from the standard cosmological model and the compatibility with the declaration parameter in different eras. Moreover, using a modified Friedmann equations we look for an expression in order to relate the ratio $r=\frac{\rho_{m}}{\rho_{ext}}$ to the Hubble parameter $H$, where $\rho_{m}$ denotes the total matter energy density (including cold dark matter contribution) and the $\rho_{ext}$ density denotes the extrinsic energy density, in order to get information of the dynamics on the ratio $r$. Finally, remarks are presented in the conclusion section.

\section{Modified Friedmann equations}

The  traditional gravitational   perturbation mechanisms in cosmology  are  essentially plagued  by coordinate  gauges, mostly
inherited  from the group of diffeomorphisms  of  general relativity. Fortunately   there  are some     very  successful
criteria to  filter out the latter perturbations \cite{Bardeen,Geroch,Walker},   but they still depend on  a  choice
of a perturbative  model.  A lesser known, but  far   more general  approach to gravitational perturbation  can be derived  from  a  theorem  due to
John   Nash,   showing   that   any  Riemannian   geometry    can be  generated by  a    continuous  sequence  of local  infinitesimal   increments of a  given  geometry \cite{Nash,Greene}.

Nash's  theorem   solves    an old  dilemma     of  Riemannian
geometry,  namely that the Riemann tensor   is not sufficient  to
make  a precise  statement    about  the local shape  of  a
geometrical object or    a manifold. The simplest   example is
given  by a 2-dimensional Riemannian  manifold,  where  the Riemann
tensor  has only  one  component $R_{1212}$  which  coincides with
the  Gaussian  curvature. Thus,   a  flat Riemannian    2-manifold
defined by  $R_{1212}=0$   may be interpreted as a plane,  a
cylinder or a   even  a helicoid,  in the sense of  Euclidean
geometry. Riemann regarded   his  concept of  curvature  as defining an  equivalent class of  manifolds instead of a specific  one \cite{Riemann}.
While  such  equivalence of  forms is  mathematically interesting,  it  is  less than  adequate     to    derive physical  conclusions
from today's  sophisticated   astronomical observations.

Using only differentiable (non-analytic) properties, Nash  showed  that any other  embedded  Riemannian geometry can  be  generated  by differentiable perturbations with a perturbed metric $\tilde{g}_{\mu\nu}= g_{\mu\nu} + \delta  g_{\mu\nu} $,  where
\be
\delta g_{\mu\nu} =-2k_{\mu\nu a}\delta y^a\;, \label{eq:York}
\ee
where  $\delta y^a$ is an infinitesimal displacement in one of the  extra dimension and $k_{\mu\nu}$ is the non-perturbed extrinsic curvature \cite{GDEI,QBW}. From this  new metric, we obtain a new extrinsic curvature $k_{\mu\nu }$  and the  procedure  can  be repeated  indefinitely:
\be
\tilde{g}_{\mu\nu}  =  g_{\mu\nu}  +  \delta y^a \, k_{\mu\nu  a}  +
\delta y^a\delta y^b\, g^{\rho\sigma}
k_{\mu\rho a}k_{\nu\sigma b}+\cdots\;,  \label{eq:pertu}
\ee
and gives the possibility to generate new geometries by smooth deformations. 

In our description of the universe, we use the Friedmann-Lema\^{\i}tre-Robertson-Walker (FLRW) line  element
in coordinates $(r,\theta,\phi,t)$ which is given by
\begin{equation}\label{eq:elementline}
ds^2=\;-dt^2+a^2\left[dr^2+f_{\kappa}^2(r)\left(d\theta^2+\sin^2\theta
d\varphi^2\right)\right]\;,
\end{equation}
where the set of functions $f_{\kappa}(r)=\sin r$, $r$,$\sinh r$ corresponds to spatial curvatures $\kappa$ = (1,
0, -1), and the function $a=a(t)$ is the expansion parameter.  This geometry can be regarded as a four-dimensional hypersurface
dynamically evolving  in  a  five-dimensional bulk space with constant
curvature. Since Nash's smooth deformations are applied to the embedding process and the FLRW geometry is completely embedded in five-dimensions \cite{rosen,szek,maia}, the coordinate $y$, usually noticed in ridig embedded models, e.g \cite{RS,RS1}, is omitted in Eq.(\ref{eq:elementline}). Concerning notation, we  use the same conventions as posed in \cite{GDEI}.

The  bulk geometry is  actually  defined by the
Einstein-Hilbert principle,   which   leads  to the Einstein
equations
\be
{\mathcal R}_{AB} -\frac{1}{2} {\mathcal R}  {\mathcal G}_{AB}=\alpha_*  T^*_{AB}\;,
\label{eq:BE0}
\ee
where $T^*_{AB}$  denotes the energy-momentum  tensor  of the known  sources. For the present application, capital Latin indices run from 1 to 5. Small case Latin indices refer to the only one extra dimension considered. All Greek indices refer to the embedded space-time counting from 1 to 4.

The  confinement  of  gauge  fields  and  ordinary  matter  are  a  standard  assumption specially in what concerns the brane-world program  as a part  of  the  solution of the hierarchy problem  of the fundamental  interactions:  the  four-dimensionality of     space-time  is  a  consequence of the invariance of Maxwell's  equations  under  the Poincar\'{e}  group. Such   condition  was  latter seen  to be  proper  to  all  gauge  fields expressed  in terms  of  differential forms  and  their  duals. However, in  spite of  many  attempts,  gravitation, in the sense of  Einstein,  does not  fit in  such  scheme.   Thus,   while  all  known  gauge  fields  are  confined to the four-dimensional   submanifold,  gravitation  as defined   in the  whole  bulk  space by  the Einstein-Hilbert principle, propagates in the  bulk.  The   proposed solution of the  hierarchy problem    says  that  gravitational  energy  scale  is    somewhere  within TeV  scale.

The  most general  expression  of this  confinement is  that  the  confined    components of  $T_{AB}$  are
proportional  to   the     energy-momentum tensor  of general relativity:  $\alpha_* T_{\mu\nu}= -8\pi G T_{\mu\nu}$. On the other  hand,   since only gravity propagates  in the  bulk   we  have  $T_{\mu a}=0$ and  $T_{ab}=0$.

Since we are dealing with embedded space-times, we need to write the induced field equations of the embedded geometry which in fact they result from the geometrical features of the bulk space by the integration of the Gauss-Codazzi equations \citep{GDEI,QBW}. To this end, we can define a five-dimensional local embedding with an embedding map $\mathcal{Z}:V_{4}\rightarrow V_{5}$. We admit that $\mathcal{Z}^{\mu}$ is a regular and differentiable map with $V_{4}$ and $V_5$ being the embedded space-time and the bulk, respectively. The components $\mathcal{Z}^{A} =\;f^{A}(x^{1},...,x^{4})$ associate with each point of $V_{4}$ a point in $V_{5}$ with coordinates $\mathcal{Z}^{A}$. These coordinates are the components of the tangent vectors of $V_{4}$. Moreover, taking the tangent, vector and scalar components of Eq.(\ref{eq:BE0}) defined in the Gaussian frame veilbein $\{\mathcal{Z}^A_{,\mu},\eta^A\}$, where $\eta^{A}$ are the components of the normal vectors of $V_{4}$, one can obtain the following equations in the embedded spacetime \citep{GDEI,QBW}
\begin{eqnarray}\label{eq:BE1}
&&R_{\mu\nu}-\frac{1}{2}Rg_{\mu\nu}-Q_{\mu\nu}=
-8\pi G T_{\mu\nu}\;, \hspace{4mm}\\
&&k_{\mu;\;\rho}^{\;\rho}-h_{,\mu} =0\;,\label{eq:BE2}
\end{eqnarray}
where  $T_{\mu\nu}$ is the four-dimensional energy-momentum tensor of
the perfect  fluid  expressed  in  co-moving coordinates as
$$T_{\mu\nu}=(p+\rho)U_{\mu}U_{\nu}+p\;g_{\mu\nu},\;\;\;U_{\mu}=\delta_{\mu}^{4}\;.$$

The quantity $Q_{\mu\nu}$  is a geometrical term defined as
\begin{equation}\label{eq:qmunu}
  Q_{\mu\nu}=g^{\rho\sigma}k_{\mu\rho }k_{\nu\sigma}- k_{\mu\nu }h -\frac{1}{2}\left(K^2-h^2\right)g_{\mu\nu}\;,
\end{equation}
where $h= g^{\mu\nu}k_{\mu\nu}$, $h^2=h.h$ and $K^{2}=k^{\mu\nu}k_{\mu\nu}$. It
follows  that the quantity  $Q_{\mu\nu}$  is  conserved in the sense  that
\begin{equation}\label{eq:cons}
  Q^{\mu\nu}{}_{;\nu}=0\;,
\end{equation}
where the symbol $(;)$ denotes the four-dimensional induced covariant derivative.

The  general  solution  for  Eq.(\ref{eq:BE2}) using the FLRW  metric is
\begin{eqnarray}
 &&k_{ij}=\frac{b}{a^2}g_{ij},\;\;
k_{44}=\frac{-1}{\dot{a}}\frac{d}{dt}\frac{b}{a}\nonumber
\end{eqnarray}
in this case $i, j= 1, 2, 3$, where  the  bending function    $b(t)=k_{11}$  is   an
arbitrary function of time, resulting from the Codazzi homogeneous equations in Eq.(\ref{eq:BE2}).

From the calculations of Eq.(\ref{eq:BE1}) and Eq.(\ref{eq:BE2}), one can obtain
\begin{eqnarray}
 &&
 k_{44}=-\frac{b}{a^{2}}\left(\frac{B}{H}-1\right)g_{44},\; h=\frac{b}{a^2}\left(\frac {B}{H}+2\right)\label{eq:hk},  \\
&&K^{2}=\frac{b^2}{a^4}\left( \frac{B^2}{H^2}-2\frac{B}{H}+4\right),\\
&&Q_{ij}= \frac{b^{2}}{a^{4}}\left( 2\frac{B}{H}-1\right)
g_{ij},\;Q_{44} = -\frac{3b^{2}}{a^{4}},
  \label{eq:Qab}\\
&&Q= -(K^2 -h^2) =\frac{6b^{2}}{a^{4}} \frac{B}{H}\;. \label{Q}
 \end{eqnarray}
In the case of Eq.(\ref{eq:Qab}), consider $i, j= 1, 2, 3$.  The usual  Hubble parameter in terms of the expansion scaling factor $a(t)=a$ is denoted by $ H = \dot{a}/a$  and the extrinsic parameter  $B= \dot{b}/b$, where the dot holds for the ordinary time derivative. It is important to point out that the determination of the bending function $b(t)$ comes from the determination of dynamical equations for extrinsic curvature. To this end, we used the ``vacuum'' Gupta equations
\begin{equation}
\label{eq:guptaflat} \mathcal{F}_{\mu\nu}=0\;,
\end{equation}
where, unlike the case of  Einstein's  equations, we do not have the equivalent to the Newtonian weak field limit and we  cannot tell about the nature of the source  term of Eq.(\ref{eq:guptaflat}) based on current experience and observations. Accordingly, the ``f-Ricci tensor'' and the ``f-Ricci scalar'', defined with $f_{\mu\nu}$ are, respectively,
$$
{\cal F}_{\mu\nu} =  f^{\alpha\lambda}\mathcal{F}_{\nu
\alpha\lambda\mu}
\;\;\mbox{and}\;\;\mathcal{F}=f^{\mu\nu}\mathcal{F}_{\mu\nu}
$$
and also the ``f-Riemann tensor''
$$
\mathcal{F}_{\nu\alpha\lambda\mu}= \;\partial_{\alpha}\Upsilon_{\mu\lambda\nu}- \;\partial_{\lambda}\Upsilon_{\mu\alpha\nu}+ \Upsilon_{\alpha\sigma\mu}\Upsilon_{\lambda\nu}^{\sigma} -\Upsilon_{\lambda\sigma\mu}\Upsilon_{\alpha\nu}^{\sigma}
$$
constructed from a  ``connection'' associated with $k_{\mu\nu}$. We stress that the geometry of the embedded space-time has been  previously   defined by $g_{\mu\nu}$. Hence, we define  the  tensors
\be
f_{\mu\nu} = \frac{2}{K}k_{\mu\nu}, \;\; \mbox{and}
\;\;f^{\mu\nu} = \frac{2}{K}k^{\mu\nu}\;,
\label{eq:fmunu}
\ee
so that $f^{\mu\rho}f_{\rho\nu} =\delta^\mu_\nu$. In the sequence we construct the ``Levi-Civita  connection'' associated with $f_{\mu\nu}$, based on the analogy with  the ``metricity condition''  $f_{\mu\nu||\rho}=0$,  where $||$ denotes  the covariant derivative with respect to $f_{\mu\nu}$ (while keeping the usual $(;)$ notation for the covariant derivative with respect to $g_{\mu\nu}$).  With this condition  we obtain the  ``f-connection''
$$
\Upsilon_{\mu\nu\sigma}=\;\frac{1}{2}\left(\partial_\mu\; f_{\sigma\nu}+ \partial_\nu\;f_{\sigma\mu} -\partial_\sigma\;f_{\mu\nu}\right)
$$
and
$$
\Upsilon_{\mu\nu}{}^{\lambda}= f^{\lambda\sigma}\;\Upsilon_{\mu\nu\sigma}\;.
$$

Replacing these results in  Eq.(\ref{eq:BE1}), we obtain the Friedman equation modified  by the extrinsic  curvature as
\begin{equation}\label{eq:Friedman}
\left(\frac{\dot{a}}{a}\right)^2+\frac{\kappa}{a^2}=\frac{4}{3}\pi
G\rho+\frac{b^2}{a^4}\;\;,
\end{equation}
where the general expression for $b(t)$ is given by
\begin{equation}\label{eq:geb}
b(t)= \alpha_0a^{\beta_0}e^{\pm
\frac{1}{2}\gamma(t)}\;,
\end{equation}
where $\alpha_0 = b_0/ a_0^{\beta_0}$ denoting $a_0$ by the present value of the expansion scaling factor and $b_0$ is an integration constant representing the present-day warp of the universe. The $\gamma$-exponent in the exponential function is given by $\gamma(t)=\sqrt{4\eta_0a^4 - 3}-\sqrt{3}\arctan\left(\frac{\sqrt{3}}{3}\sqrt{4\eta_0a^4 -3}\right)$. The two signs represent two possible signatures of the evolution of the bending function $b(t)$ in which we denote for simplicity $\gamma^{+}$ and $\gamma^{-}$ solutions. The $\beta_0$ parameter affects the magnitude of the deceleration parameter $q$ and the $\eta_0$ parameter measures the width of the transition phase $z_t$ from a decelerating to accelerating regime. These two parameters were generated essentially from Eqs.(\ref{eq:cons}) and (\ref{eq:guptaflat}), respectively, as shown in \cite{GDEI}.

Accordingly, using Eqs.(\ref{eq:Friedman}) and (\ref{eq:geb}) we can write Friedmann equations in a form
\begin{equation}\label{eq:2}
H(z)=H_0\sqrt{\Omega_{\;m}(1+z)^3 +
\Omega_{\;ext}(1+z)^{4-2\beta_0}e^{\pm\gamma(z)}}\;,
\end{equation}
where $H(z)$ is the Hubble parameter in terms of redshift $z$ and $H_0$ is the current Hubble constant. The matter density parameter is denoted by $\Omega_m$ and the term $\Omega_{ext}$ stands for the density parameter associated with the extrinsic curvature.

\section{Alleviating the coincidence}
The coincidence problem basically results in the explanation of the why matter density contribution $\rho_m$ and the vacuum contribution $\rho_{\Lambda}$ are about the same order at present time. The \emph{alleviation} occurs when the contributions $\Omega_m/\Omega_{\Lambda} < (1+z)^3$ \citep{dalal,alc} and is used to select cosmological models. Since we are not attributing any dynamical property to CC, we use the extrinsic curvature to make the appropriate correction adding an extra-information to this framework with its ``extrinsic'' contribution $\Omega_{ext}$. Moreover, using ``fluid analogy'' and Eq.(\ref{eq:Friedman}) we denote $\Omega_{ext}$ as
$$\Omega_{ext}= \frac{b^2}{a^4}= \frac{8\pi G }{3}\rho_{ext}\;,$$
where $\rho^0_{ext}$ denotes the current extrinsic energy density. It is important to notice that matter energy density and extrinsic energy density are conserved independently in the sense that $T_{\mu\nu;\nu}=0$ and, according to Eq.(\ref{eq:cons}), $Q_{\mu\nu;\nu}=0$. Thus, we can write the conservation equation for matter as
$$\dot{\rho}_m + 3H \rho_m =0\;,$$
where the dot symbol denotes the time derivative. Moreover, using Eq.(\ref{eq:Friedman}), we denote
$$\rho_{ext} = \rho_{ext}^0 (1+z)^{4-2\beta_0}\;,$$
where $\rho_{ext}^0$ is the current extrinsic energy density, one can get the conservation equation for  the extrinsic contribution as
$$\dot{\rho}_{ext} + (4-2\beta_0)H \rho_{ext} =0\;.$$

At first, due to its dynamical characteristics, we regard the extrinsic curvature as the main cause of the current accelerated expansion rather than CC in accordance with \cite{GDEI}. Since they are two different quantities, the coincidence tends to vanish once the dynamics of extrinsic curvature can be understood. To this end, we define the ratio $r=\frac{\rho_{m}}{\rho_{ext}}$ and study its behavior seeking a relation with $r$ to the Hubble parameter as we do not have an independent observational data of the ratio $r$. As a starting point, we adopt the current value as $r_0=3/7$ \citep{delcampo} as an input parameter. Bearing in mind that the modified Friedmann equations as shown in Eq.(\ref{eq:Friedman}) they can be written in terms of energy density as
$$3H^2= \kappa^2 \left(\rho_{m} + \rho_{ext} \exp{(\pm \gamma)}\right)\;,$$
and one can obtain
\begin{equation}\label{eq:rdot}
\dot{r}= (1-2\beta_0)r H \;.
\end{equation}
Moreover, from the analysis of the critical point from the direct derivation of Eq.(\ref{eq:rdot}), one can obtain
\begin{equation}\label{eq:rdot2}
\left(\frac{\dot{r}}{r}\right)\rfloor_0= (1-2\beta_0)^2r H^2_0 > 0\;.
\end{equation}
As it happens, the minima points occur at $\beta_0 < 1/2$, which means that $r$ is bounded from below.

In addition, Eq.(\ref{eq:rdot}) can be easily integrated and gives the solution
\begin{equation}\label{eq:rz}
r(z)= r_0 (1+z)^{\left(2\beta_0 -1\right)}\;,
\end{equation}
in terms of redshift, where we have used the relation $\frac{d}{dt} =- H(1+z) \frac{d}{dz}$.

It is important to notice that the pair of parameters $(\beta_0, \eta_0)$ in Eq.(\ref{eq:Friedman}) was already constrained in the model presented in \citep{Capistrano}. For instance, we adopt the values $\beta_0=2$, $\eta_0=0.001$ and $\eta_0=0.25$ that passed through cosmokinetics tests. For accelerated expansion it was shown that $2\leq \beta_0 \leq 3$ \cite{Capistrano}. Moreover, the solution with $\beta_0=2$ and $\eta_0=0.25$ can lead to a recollapsing universe with a total density parameter $\Omega > 1$. Based on the fact that Eq.(\ref{eq:Friedman}) can provide different solutions with the term $\gamma$, for the special case when $\pm \gamma(z) = 0$, one can obtain XCDM-like patterns as shown in \cite{GDEI} with a correspondence
\begin{equation}\label{eq:exotic}
4-2\beta_0 = 3(1+ w)\;,
\end{equation}
where $w$ is a dimensionless parameter of the X-fluid equation of state $w=\frac{p}{\rho}$ \citep{turner} as a ratio between its pressure $p$ and density $\rho$. Moreover, we adopt the current value of Hubble constant $H_0$ as $H_0=67.8 \pm 0.9\;km.s^{-1}.Mpc^{-1}$ based on the latest observations \citep{Ade}.

\begin{figure}
  \includegraphics[width=3.5in, height=4in]{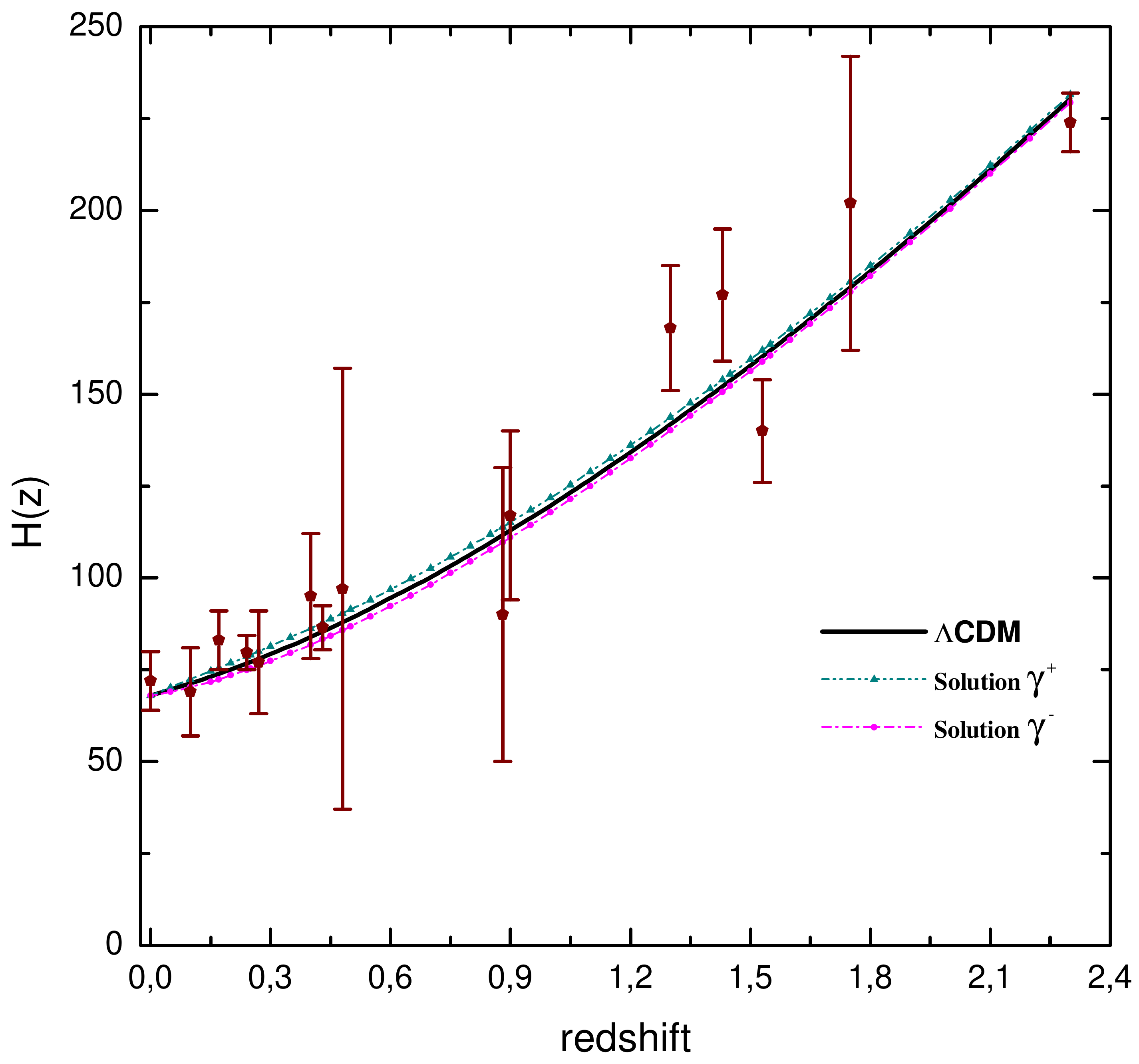}\\
 \caption{As an example, we show the Hubble parameter as a function of redshift for two models \cite{Capistrano}. The model I is defined by  the $\gamma^{(+)}$ and $\gamma^{(-)}$ solutions with the values for $\beta_0=2$ and $\eta_0=0.001$, the curves coincide with the curve from $\Lambda$CDM (solid line). Moreover, for the model II  the curves are shown with $\gamma^{(+)}$ (triangles) and $\gamma^{(-)}$ (circles) solutions for the values $\beta_0=2$ and $\eta_0=0.25$. Error bar points were extracted from \citep{cao}  supplemented with additional data from \citep{busca,zhai} at $H(z = 2.3)$.}\label{fig:hubposneg}
\end{figure}

In Fig.(\ref{fig:hubposneg}), we have used Eq.(\ref{eq:2}) and present the evolution of the Hubble parameter for the two models ($\gamma^{(+)}$, $\gamma^{(-)}$) with adopted values of $(\beta_0, \eta_0)$. The graph also shows that those solutions are very close to $\Lambda$CDM prediction (solid line).

In order to do not jeopardize the nucleosynthesis, we use $|r_{nuc}=r(z \sim 10^9)|\leqslant 10\% $ \cite{grande,sola} that gives a constraint $\beta_0\lesssim 0.465$. Moreover, from Eq.(\ref{eq:2}), we can write the deceleration parameter conveniently written in terms of the redshift $z$ as
\begin{equation}\label{eq:q(z)}
q(z) =\frac{1}{H(z)}\frac{dH(z)}{dz}(1 + z) - 1\;.
\end{equation}
Hence, we can write
\begin{equation}\label{eq:q(z)2}
q(z)= \frac{3}{2}\left[\frac{\Omega_{\;m}(1+z)^3 + \gamma^{*}
\Omega_{\;ext}(1+z)^{4-2\beta_0}e^{\mp \gamma(z)}}
{\Omega_{\;m}(1+z)^3+\Omega_{\;ext}(1+z)^{4-2\beta_0}e^{\mp\gamma(z)}}\right]-1\;\;,
\end{equation}
where $\gamma^{*}=\frac{1}{3}\left[4-2\beta_0 \pm 2 \sqrt{\frac{4\eta_0}{(1+z)^4} - 3}\;\right]$.

Using the related value of $\beta_0$ for nucleosynthesis era, the baryon contribution $\Omega_b:=0.022\pm 0.00023$ and Cold dark matter contribution $0.1197\pm0.0022$ with $68\%$C.L. \cite{Ade}, we obtain the deceleration parameter $q\sim 0.535$ with the expected ratio $r\sim 0.1$. Moreover, for $\beta_0 = 1/2$ (that corresponds to the matter dominated era with $w=0$ in accordance with Eq.(\ref{eq:exotic}) ) and $\eta_0=0.001$, we obtain the predicted value $q=1/2$ expected for the matter domination and putting  $\beta_0 = 1/2$ in Eq.(\ref{eq:rz}) it converges to the value $r_0=3/7$. Actually, the same value for the deceleration parameter is obtained even considering the highest value of $\eta_0=0.5$. Moreover, taking Eq.(\ref{eq:q(z)2}) with the previous values for $\beta_0=1/2$ and $q = 1/2$ related to matter-dominated era, we can obtain an estimative for the magnitude of the ``equivalence redshift'' $z_e$ given by
\begin{equation}\label{eq:redc}
z_e=|\left(\frac{4}{3}|\eta_0|\right)^{(1/4)} -1|\;.
\end{equation}

\begin{figure*}
\subfloat[]{%
  \includegraphics[height=6cm,width=.49\linewidth]{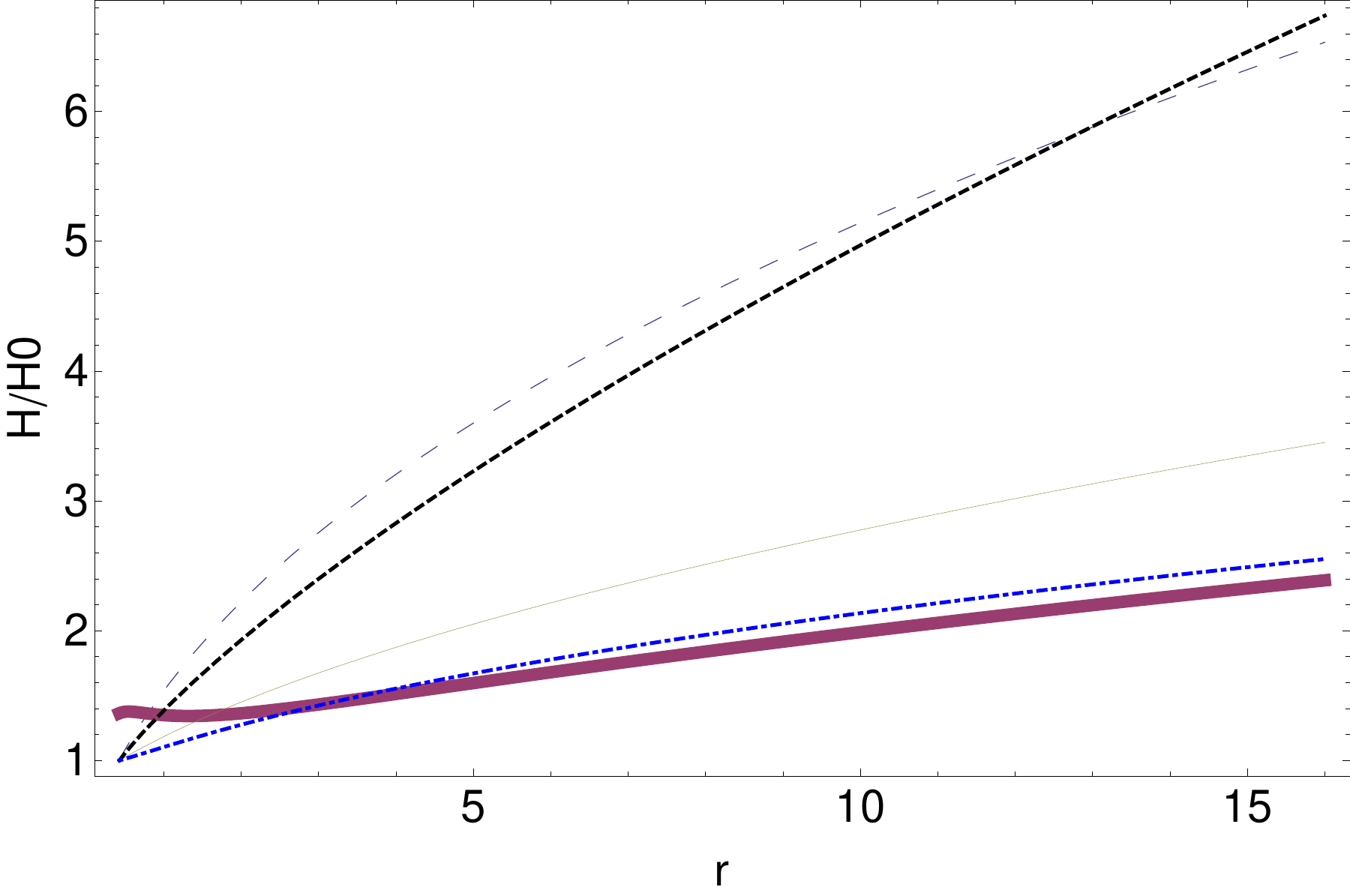}%
}\hfill
\subfloat[]{%
  \includegraphics[height=6cm,width=.49\linewidth]{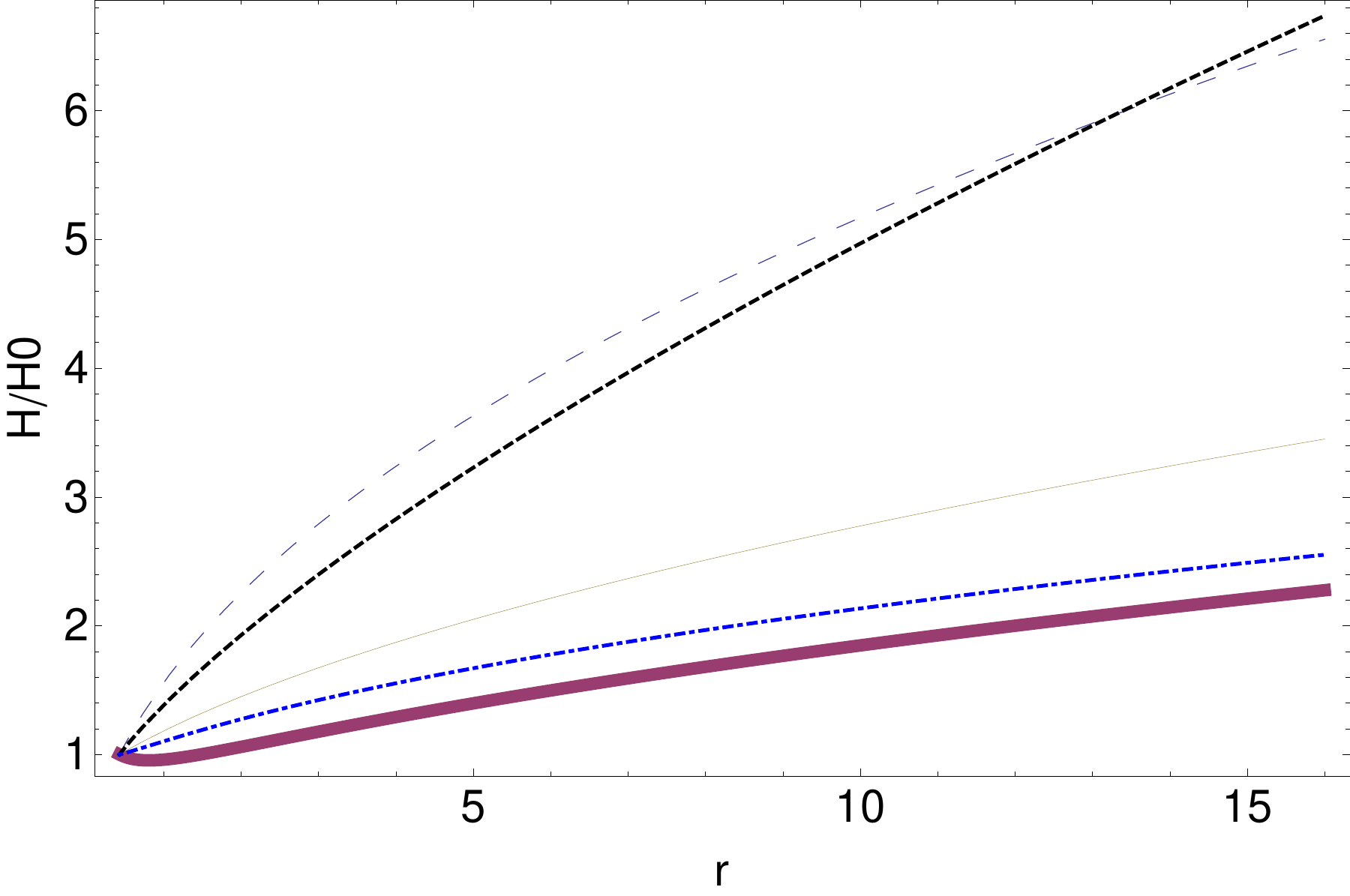}%
}
\caption{The relation of Hubble parameter to ratio $r$ raging from  the input parameter $r_0=3/7$ to $r=15.4$ where is present the solutions of the model, including a XCDM-like solutions for $\pm \gamma=0$. The thin line mimics the $\Lambda$CDM model with $w=-1$, the thick dashed line mimics a quintessence model with $w=-0.73$. Phantom models with $w=-1.2$ are represented by the thick dotted-dashed line. The dashed lines and thick lines are solutions for $\pm\gamma\neq 0$.} \label{fig:r}
\end{figure*}

In order to approximate the matter dominated era we know that the variation of the $\eta_0$ parameter has a constrained small value in the accelerated expansion where was found that $0\leq \eta_0 \leq 0.5$ that gives the range $ 1 \geq z_e \geq 0.09$. Interestingly, if we consider a tighter range for the width of transition, i.e., $0.01\leq \eta_0 \leq 0.25$, we have $ 0.66 \geq z_e \geq 0.24$ which means that the beginning of the ``coincidence'' happens in the end of the matter dominated era, since the value of $z_e$ matches the transition redshift from a deceleration to accelerated phase, as several different data sets indicate, e.g, $z_t = 0.56^{+0.13}_{-0.10}$ as the latest Baryons Acoustic Oscillations/Cosmic Microwave Background (BAO/CMBR) + SNIa \cite{Giostri} with MLCS2K2 light-curve fitter or using the SALT2 fitter $z_t = 0.64^{+0.13}_{-0.07}$. This leads us to the interesting conclusion that the apparent ``coincidence'' began in the matter-dominated era and the current ``coincidence'' observed is far from being special.

It is important to point out that the apparent variation of the $\beta_0$ parameter is from its inner relation to the deceleration parameter $q$ and any change of $\beta_0$ is related to phase transitions in the universe. As previously commented, the current accelerated expansion (at $z=0$) regime can be obtained with $2\leq\beta_0\leq 3$ and with the expected $q_0<0$. In short, based on the fact that we have minima points such that $\beta_0<1/2$, the $\beta_0$ parameter can be constrained as $0.465 \leq \beta_0 \leq 3$ from nucleosynthesis until the present-day accelerated expansion regime.

In addition, we try to obtain more information on the evolving $r$ looking for a relation between Hubble parameter and the ratio $r$. At first, one can write the following relation
\begin{equation}\label{eq:relr}
\dot{r}=  \dot{H} \frac{dr}{dH}\;.
\end{equation}
Starting from calculating the first derivative of the Friedman equation in Eq.(\ref{eq:Friedman}) in terms of densities $\rho_{m}$ and $\rho_{ext}$, one can obtain
$$\dot{H} = \frac{\kappa^2}{6}\left( (2\beta_0-4) \rho_{ext}\exp{(\pm \gamma)} - 3\rho_m \pm \rho_{ext}\dot{\gamma} \frac{\exp{(\pm \gamma)}}{H} \right) \; .$$

After a long algebra, one can write
$$\dot{H} = -\frac{3}{2} \left[ \frac{r+ \Theta\exp{(\pm \gamma)}}{r + \exp{(\pm \gamma)}}\right]  H^2 \;.$$
Interestingly, if we consider a particular case (i.e, to mimic XCDM), one can set the term $\gamma=0 $ and the appropriate correspondence $\beta_0 = \frac{1}{2}\left(1-3w\right)$, where $w < -1/3$. So we have $\Theta= 1+w$. Hence,
$$\dot{H} = -\frac{3}{2} \left[ \frac{1+ w + r}{r + 1}\right]  H^2 \;,$$
which is the same equation obtained in \citep{delcampo}.

Moreover, the relation in Eq.(\ref{eq:relr}) can be readily integrated and after a long algebra, one can write
$$H = H_0 \left[ \frac{r+ \exp{(\pm \gamma)}}{r_0 + \exp{(\pm \gamma)}}\right]^{-\frac{3}{2}\frac{1-\Theta}{(1-2\beta_0)}} \left(\frac{r}{r_0}\right)^{-\frac{3}{2}\frac{(\Theta)}{(1-2\beta_0)}}\;,$$
where the Hubble parameter $H$ is a function of $r$. The function $\Theta(r)$ is denoted by
$$\Theta(r)= \frac{1}{3}\left[ (4-2\beta_0) \pm 2 \sqrt{ 4 \eta_0 \left( \frac{r}{r_0}\right)^{4/(2\beta_0 -1)}- 3} \right]\;, $$
with two signs $\Theta^{+}$ and $\Theta^{-}$. These two possibilities induce to four possible behaviors of $H(r)$  which we denote $H(r)^{+-}, H(r)^{-+}, H(r)^{++}, H(r)^{--}$ as shown below.
\begin{eqnarray*}
  H(r)^{+-} &=& H_0 \left[ \frac{r+ \exp{(+ \gamma)}}{r_0 + \exp{(+ \gamma)}}\right]^{-\frac{3}{2}\frac{1-\Theta^{-}}{(1-2\beta_0)}} \left(\frac{r}{r_0}\right)^{-\frac{3}{2}\frac{(\Theta^{-})}{(1-2\beta_0)}}\; \\
  H(r)^{--} &=& H_0 \left[ \frac{r+ \exp{(- \gamma)}}{r_0 + \exp{(- \gamma)}}\right]^{-\frac{3}{2}\frac{1-\Theta^{-}}{(1-2\beta_0)}} \left(\frac{r}{r_0}\right)^{-\frac{3}{2}\frac{(\Theta^{-})}{(1-2\beta_0)}}\; \\
  H(r)^{++} &=& H_0 \left[ \frac{r+ \exp{(+ \gamma)}}{r_0 + \exp{(+ \gamma)}}\right]^{-\frac{3}{2}\frac{1-\Theta^{+}}{(1-2\beta_0)}} \left(\frac{r}{r_0}\right)^{-\frac{3}{2}\frac{(\Theta^{+})}{(1-2\beta_0)}}\; \\
  H(r)^{-+} &=& H_0 \left[ \frac{r+ \exp{(- \gamma)}}{r_0 + \exp{(- \gamma)}}\right]^{-\frac{3}{2}\frac{1-\Theta^{+}}{(1-2\beta_0)}} \left(\frac{r}{r_0}\right)^{-\frac{3}{2}\frac{(\Theta^{+})}{(1-2\beta_0)}}\;
\end{eqnarray*}

In Fig.(\ref{fig:r}) we present some results of different models with an evolving Hubble parameter as the ratio $r$ increases (and the coincidence ceases to be). As a particular solution of our model, one can get the curves in fig.(\ref{fig:r}) that mimics XCDM, as previously commented. The thin line mimics the $\Lambda$CDM model with $w=-1$, the thick dashed line mimics a quintessence model with $w=-0.73$. Phantom models with $w=-1.2$ are represented by the thick dotted-dashed line. In the left panel, the dashed line and the thick line represent the functions $H(r)^{++}$ and $H(r)^{+-}$, respectively. In the right panel, the dashed line and the thick line represent the functions $H(r)^{-+}$ and $H(r)^{--}$, respectively.

We note that both left and right panels present a general similar behavior. Starting with $H(r)^{++}$ and $H(r)^{-+}$ one finds curves very close to a quintessence model. Moreover, the $H(r)^{+-}$ and $H(r)^{--}$ are close to phantom. The solution $H(r)^{+-}$ presents a smooth decaying at $r<1$ differently from $H(r)^{--}$ that mimics a more negative equation of state with $w=-1.3$ . It is worth noting to say that the evolution of $H(z)$ in terms of redshift was shown in \citep{Capistrano} and the values of parameters $(\eta_0=0.25,\beta_0=2)$ passed through cosmokinetics tests confronted to observational data from \citep{cao} based on observations of red-enveloped galaxies \citep{stern} and BAO peaks \citep{gaztanaga} being also supplemented with the observational data on Hubble parameter (OHD) and BAO in Ly$\alpha$ \citep{busca,zhai} with $H(z = 2.3) = (224 \pm 8) km.s^{-1}.Mpc^{-1}$. As it may seem, the apparent coincidence is caused by the effect of extrinsic curvature on the dynamics of the universe mainly on the variation of the deceleration parameter.

\section{Final remarks}
Until   recently  the problem of classes  of  equivalence of  manifolds  defined by the same  Riemann curvature,  but  with  different topological properties, was  ignored in  general  relativity because  the  Minkowski  space-time,  postulated   as  the ground  state  of  gravitational  field,   is uniquely  and  well defined  as  a  flat-plane  manifold,  characterized by the existence of Poincar\'e  translations in all directions  in all of its points.  However,   recent  astrophysical  experiments such  as, e.g, WMAP, SDSS and Planck Mission, have  shown  that exists a small  CC   which  explains  the accelerated  expansion  within the  so  called  $\Lambda$CDM  paradigm.  In this case,  we  no longer  have  the Minkowski  space-time  as  a  solution of Einstein's  equations, so that  the  ground  state   of the  gravitational field, or  equivalently  the   Minkowski standard  of Riemann curvature  set by  Einstein  is  ambiguous: Either  we have  Minkowski without  CC or  else  we  have  de Sitter  space-times  with CC. In order to solve that problem, the embedding between geometries has been revealed to be an appropriate underlying mechanism to obtain a gravitational theory and a possible way to get to a quantum-gravity theory \cite{QBW, rostami, jalal} in the future. The  relevant  detail in  Nash's theorem  is that it provides   a  mathematically sound and coordinate gauge free way to construct any Riemannian  geometry, and  in particular any space-time  structures by a  continuous sequence of infinitesimal perturbations along the  extra dimensions of the  bulk space generated by the  extrinsic  curvature. Due to its perturbative characteristics, Nash's  geometric smooth perturbative process can be an important principle to cosmological applications.

We focused our present work on the ``new'' CC problem commonly referred as coincidence problem on the values of the contributions of matter and the vacuum and the why they seem to have the same order. Motivated by the former geometrical dilemma on Riemann's geometry, we studied the contribution of the extrinsic curvature as a dynamical quantity and its influence to the coincidence problem as we replace the CC contribution by the extrinsic one. It was shown that the coincidence problem can be at least alleviated with the presence of the extrinsic curvature. Interestingly, the $\beta_0$ parameter revealed to be very promising term since it is related to the magnitude of the declaration parameter $q$ and its small changes are related to the transition phases of the universe. It helped us to understand that the current coincidence ceases to be since it began close to the end of the matter dominated or even in the passage from the decelerated to accelerated regime, where we found the ``equivalence'' redshift around $0.66 \geq z_e \geq 0.24$ compatible with the transition phase. Moreover, the parameter $\beta_0$ could be highly constrained as $0.465 \leq \beta_0 \leq 3$ from nucleosynthesis until the present-day accelerated expansion era showing how the ratio $r$ evolved to its present value $r_0=3/7$. An explicit relation of the ratio of the densities $r= \frac{\rho_{m}}{\rho_{ext}}$ to the Hubble parameter was obtained in order to understand more the dynamics on the ratio $r$. This analysis can be improved as the observational data of the ratio $r$ can be available in future observations.

\end{document}